\documentclass[aps,prl,superscriptaddress,amsmath,amssymb,floatfix,twocolumn,showpacs]{revtex4-1}
\usepackage{times}
\usepackage{latexsym,amsmath,amssymb,bm,euscript}
\usepackage{bbm}
\usepackage{multirow}
\usepackage{dcolumn}
\usepackage{wasysym}
\usepackage{graphicx}
\usepackage{graphics}
\usepackage{subfigure}
\usepackage{color}
\usepackage{epstopdf}
\usepackage{slashed}
\usepackage{bm}
\usepackage{mathtools}
\usepackage{braket}
\usepackage{soul}

\def\ra{\rangle}
\def\la{\langle}

\def\Hc{{\rm H.c.}}

\def\rhombpic{{\begin{picture}(26,15)(-2,-2)
                     \put (6,0) {\circle*{5}}
		     \put (18,0) {\circle*{5}}
		     \put (6,10) {\circle*{5}}
		     \put (18,10) {\circle*{5}}
		     \put (6,10) {\line (1,0) {12}}
		     \put (6,0) {\line (1,0) {12}}
                     \put (18,10){\line(0,-1){12}}
		     \put (6,0) {\line (0,1) {12}}
               \end{picture}}}

\def\triangpic{{\begin{picture}(17,15)(-2,-2)
                      \put (0,0) {\circle*{5}}
		      \put (12,0) {\circle*{5}}
		      \put (6,10) {\circle*{5}}
		      \put (0,0) {\line (1,0) {12}}
		      \put (12,0) {\line (-3,5) {6}}
		      \put (0,0) {\line (3,5) {6}}
                \end{picture}}}

\begin{document}

\title{Frustrated magnetism and bicollinear antiferromagnetic order in FeTe}

\author{Hsin-Hua Lai}
\affiliation{Department of Physics and Astronomy \& Rice Center for Quantum Materials, Rice University, Houston, Texas 77005, USA}
\author{Shou-Shu Gong}
\affiliation{National High Magnetic Field Laboratory, Florida State University, Tallahassee, Florida 32310, USA}
\author{Wen-Jun Hu}
\affiliation{Department of Physics and Astronomy \& Rice Center for Quantum Materials, Rice University, Houston, Texas 77005, USA}
\author{Qimiao Si}
\affiliation{Department of Physics and Astronomy \& Rice Center for Quantum Materials, Rice University, Houston, Texas 77005, USA}

\begin{abstract}
Iron chalcogenides display a rich variety of electronic orders in their phase diagram. A particularly enigmatic case is FeTe, a metal which possesses co-existing hole and electron Fermi surfaces as in the iron pnictides but has a distinct ($\pi$/2,$\pi/2$) bicollinear antiferromagnetic order in the Fe square lattice. 
While local-moment physics has been recognized as essential for understanding the electronic order, it has been a long-standing challenge to understand how the bicollinear antiferromagnetic ground state emerges in a proper quantum spin model. We show here that a bilinear-biquadratic spin-$1$ model on a square lattice with nonzero ring-exchange interactions exhibits the bicollinear antiferromagnetic order over an extended parameter space in its phase diagram. Our work shows that frustrated magnetism in the quantum spin model provides a unified description of the electronic orders in the iron chalcogenides and iron pnictides.
\end{abstract}

\pacs{74.70.Xa,75.10.Jm,71.10.Hf, 71.27.+a}

\maketitle
\textit{Introduction.}---
Iron-based superconductors (FeSCs) have been of extensive interest during the past eight years \cite{Kamihara2008, Johnston2010, PCDai2015, Si2016}. Early work in the field focused on the iron pncitides, such as BaFe$_2$As$_2$ with various chemical substitutions. More recently, the iron chalcogenides have occupied the center stage, in part because they have provided a new record of the superconducting transition temperature ($T_c$) \cite{QYWang2012,JJLee_Nature,SLHe2013,YWang2015} and a renewed hope of reaching even higher $T_c$. Because superconductivity in these materials occurs at the border of correlation-induced electronic orders \cite{Johnston2010, Si2016}, it is vitally important to understand the origin and nature of the ordered states. 

\begin{figure}[t]
    \subfigure[]{\label{Fig:bicollinear}\includegraphics[width= 1.08 in]{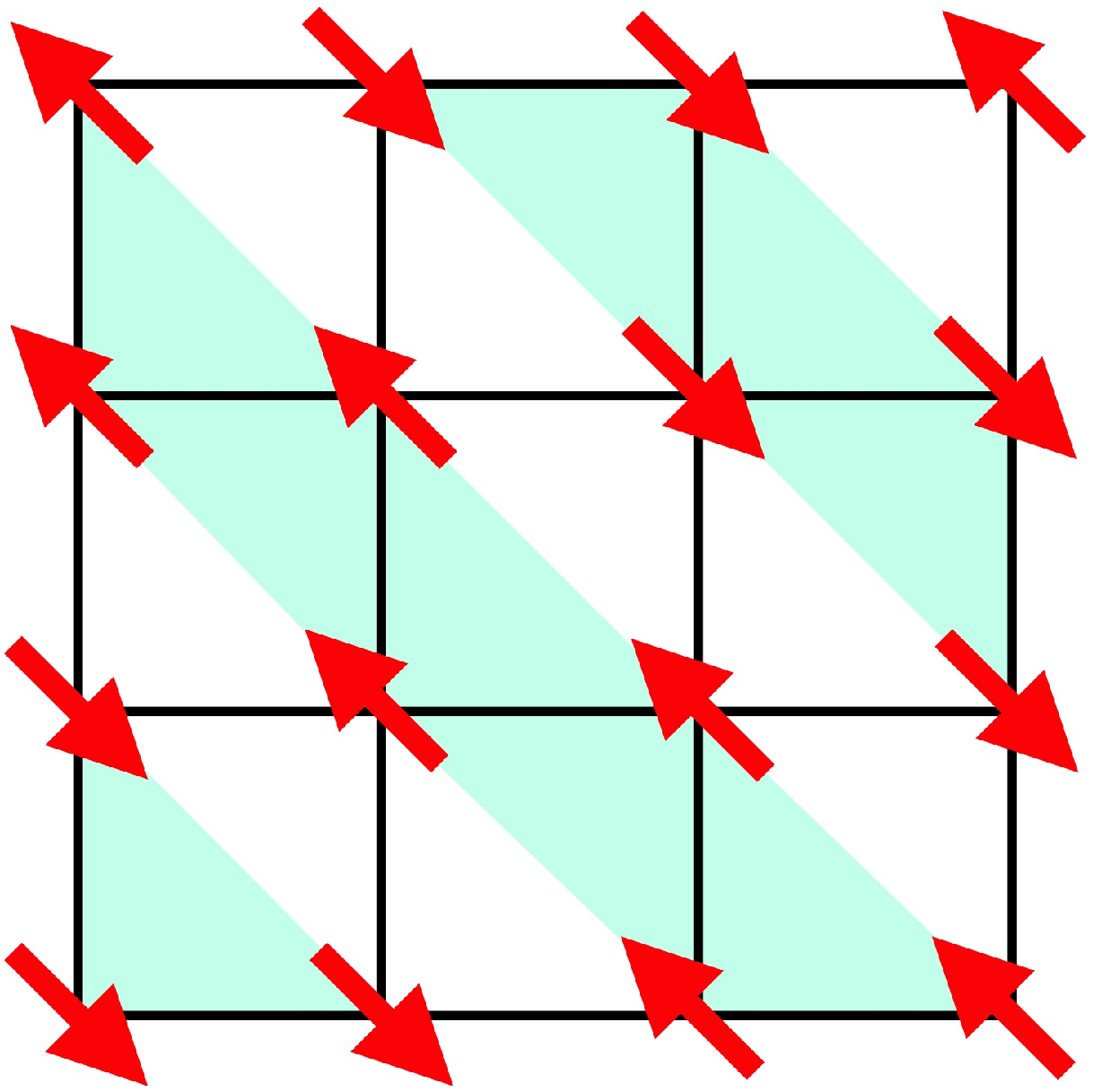}}
    \subfigure[]{\label{Fig:plaquette}\includegraphics[width=1 in]{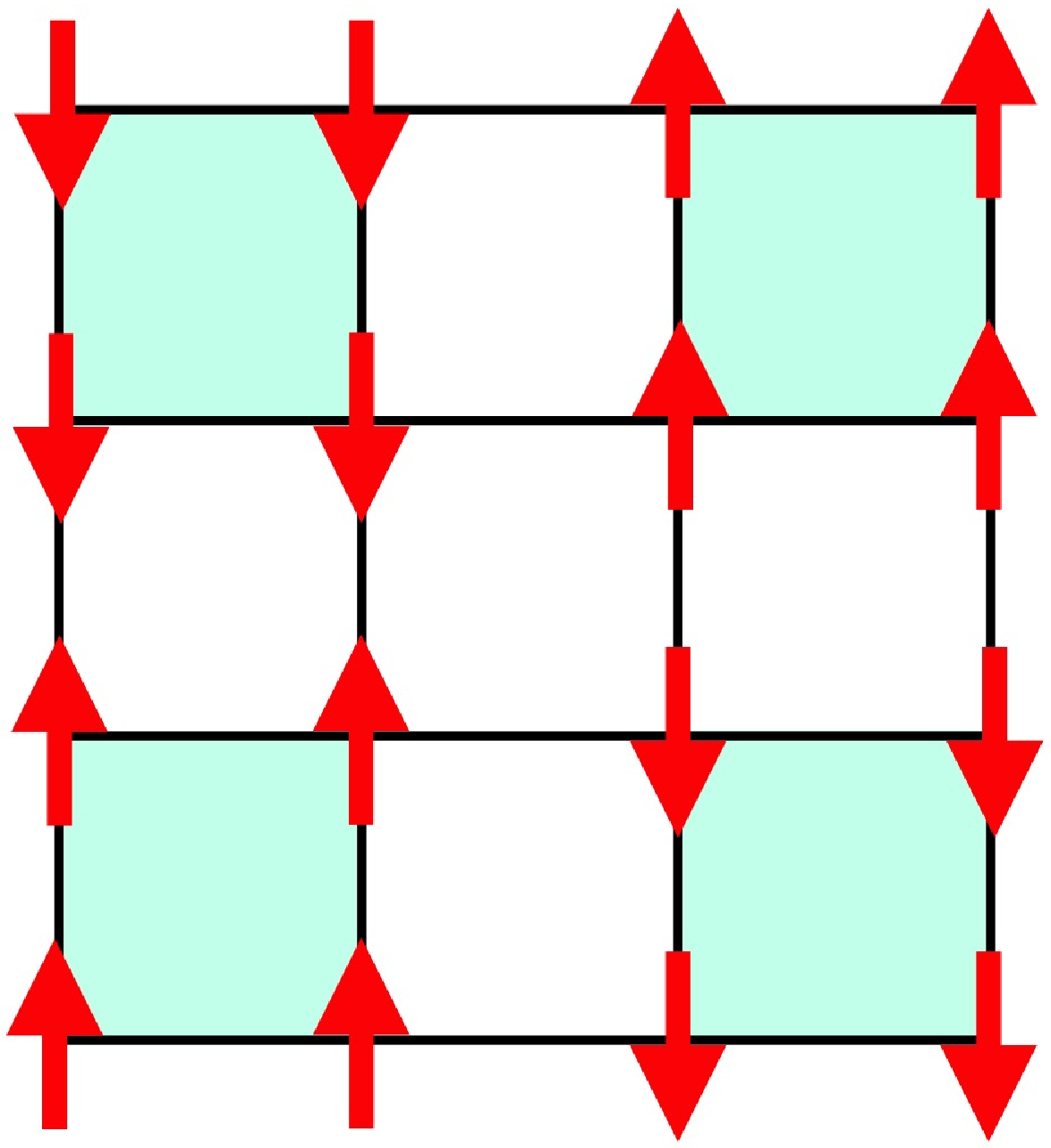}}\\
     \subfigure[]{\label{Fig:AFMs_a}\includegraphics[width= 1.05 in]{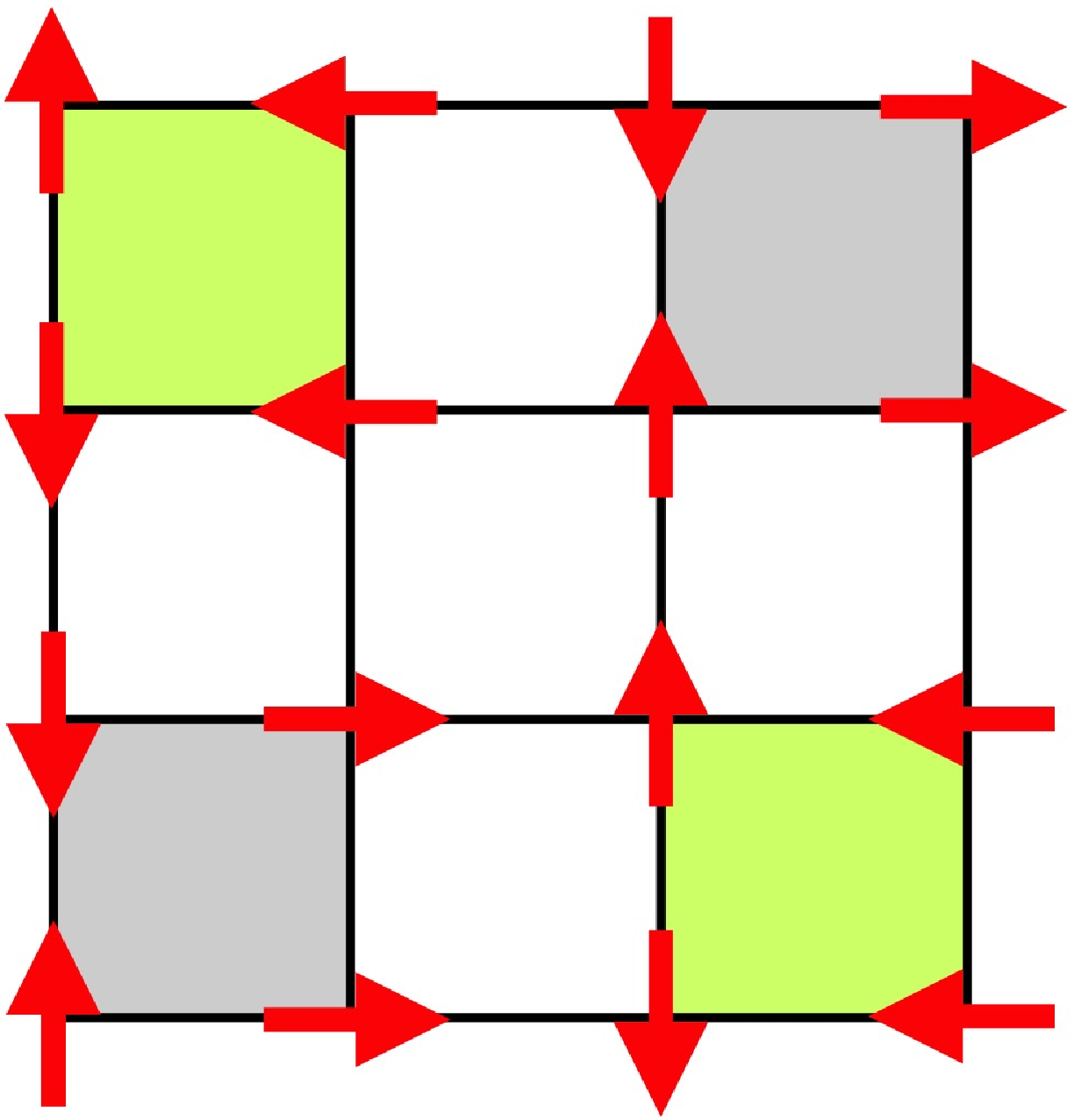}}
    \subfigure[]{\label{Fig:AFMs_b}\includegraphics[width=1.05 in]{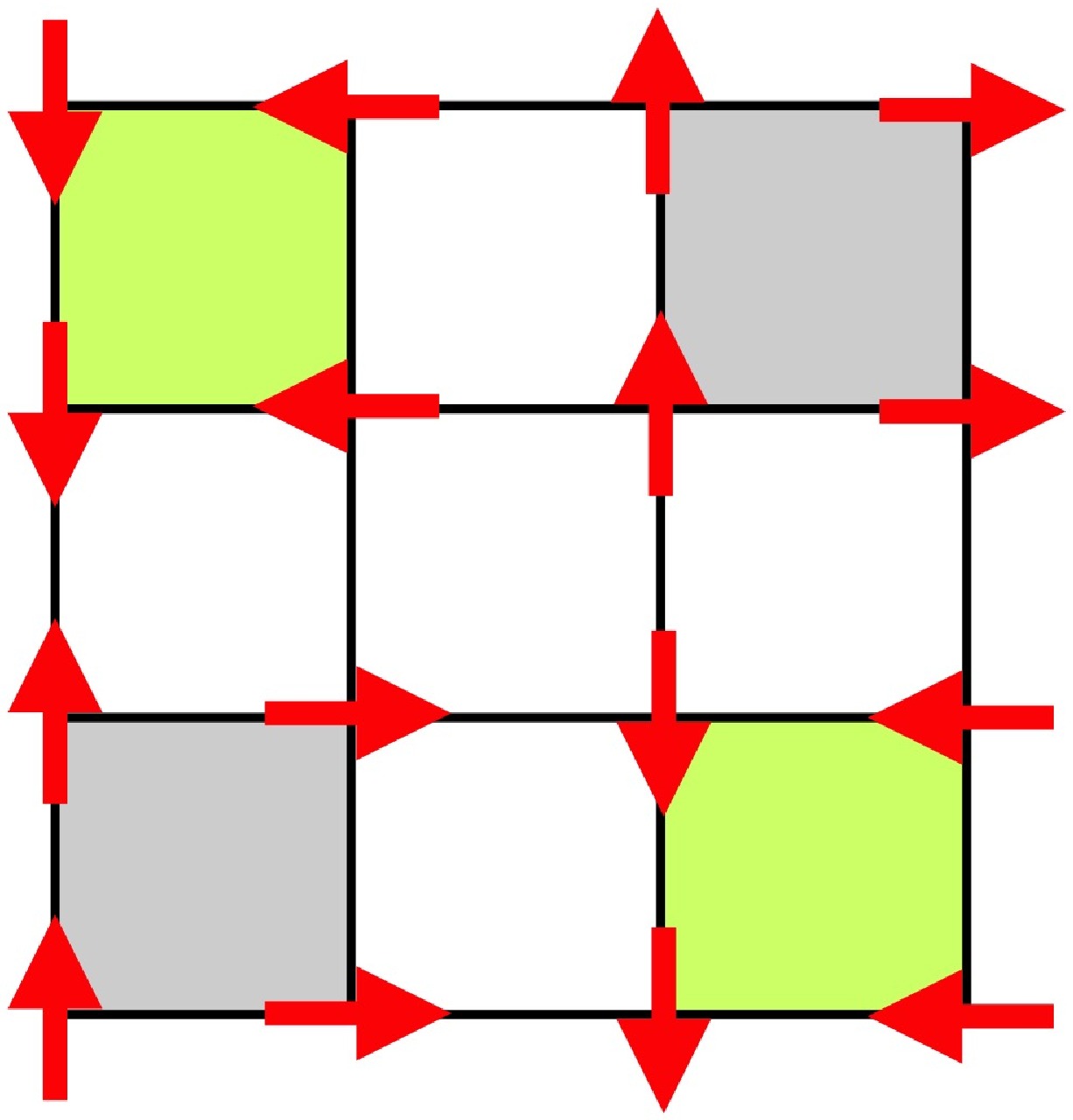}}
    \caption{(color online) Spin configuration of each magnetic order state. (a) and (b) are the $(\pi/2,\pi/2)$ bicollinear and $(\pi/2, \pi/2)$ plaquette AFM orders. The light blue regimes illustrate the double stripe and plaquette patterns in (a) and (b). These two orders remain degenerate in the $J_n$-$K_n$ ($n=1,2,3$) bilinear-biquadratic interactions. (c) and (d) are the  $(\pi/2,\pi/2)$ AFM$_a^*$ and $(\pi/2,\pi/2)$ AFM$_b^*$, respectively. The two AFM$^*_{a/b}$ orders are also degenerate in the $J$-$K$ model.}
\label{Fig:pi/2pi/2_phase}
\end{figure}

One of the outstanding puzzles in the field arises in the structurally simplest iron chalcogenide FeTe. Compared to the iron pnictides, it has a similar Fermi surface with hole and electron pockets \cite{Subedi2008,Xia2009}. Yet, instead of having a ($\pi$, $0$) collinear antiferromagnetic (AFM) order as in the latter case, FeTe has a ($\pi/2$,$\pi/2$) bicollinear AFM order (BC) illustrated in Fig.~\ref{Fig:bicollinear} \cite{Bao2009,ShiliangLi2009,JWen_FeTe}. This order is entirely unexpected in the weak-coupling Fermi-surface nesting picture. Thus, it is widely believed that
the origin lies in the frustrated magnetism of correlation-induced local moments \cite{SiAbrahams08, Ma2009, Fang2009}. 
A natural starting point would be to consider bilinear exchange interactions between the local moments with nearest-neighbor ($J_1$) and further neighbor (second-neighbor $J_2$ and third-neighbor $J_3$) interactions on the square lattice. However, a classical spin model with bilinear $J_1-J_2-J_3$ interactions yields incommensurate spiral magnetic order, and the ordering wavevector can reach $(\pi/2, \pi/2)$ only for an infinitesimal $J_1$ \cite{Moreo1990, Ferrer1993}. 
The multi-band nature of these systems makes it natural to consider the role of the biquadratic interactions,
which have recently been discussed
\cite{YuSi_AFQ,FaWang2015,Glasbrenner2015, Wangzhentao2016,Lai2016, Wenjun_nematic2016,Ergueta2016} 
as providing a route towards understanding the intriguing phenomenologies of the iron chalcogenides.
Indeed, it is also known in classical spin models that the presence of the biquadratic $K$ interactions can make 
the ordering 
wavevector to be $(\pi/2,\pi/2)$ \cite{Hu2012}. The problem is that the BC state is degenerate with the plaquette 
AFM order (PL) shown in Fig.~\ref{Fig:plaquette}. How the BC order emerges as the true ground state 
in quantum spin models remains a long-standing puzzle.

In this \textit{Letter}, we propose that ring exchange ($\mathfrak{R}$) interactions provide a robust mechanism to stabilize the BC order. Strongly correlated bad metals such as FeTe possess significant charge fluctuations \cite{Si2016}. We therefore expect that the ring exchange interactions involving cyclic permuting spin degrees of freedoms on more-than-two lattice sites can be significant, in addition to the biquadratic terms that only capture charge fluctuations between two lattice sites.

More specifically, we study the frustrated quantum spin model on a square lattice that contains bilinear and biquadratic interactions. For spin $S=1$, using semi-classical site-factorized wavefunction analysis and fully-quantum density matrix renormalization group (DMRG) studies, we provide evidence for a phase regime in which the BC order is degenerate with the PL order. The degeneracy is robust, persisting even for the higher $S=3/2, 2$ systems as found by the DMRG calculations. We then determine the fluctuation spectra of the BC and PL phases using a flavor-wave theory analysis for the spin-$1$ case. This allows us to identify a regime where quantum fluctuations select the BC order, by destabilizing the PL order. For the much more extended parameter regime where both the BC and PL states are stable and degenerate, we show that a nonzero $4$-site ring-exchange interaction selects the BC order. 
 
\textit{Model.}--- 
We study the $J$-$K$-$\mathfrak{R}$ model whose Hamiltonian is defined as
\begin{eqnarray}
 H&=&  \sum_{i, j} \left[ J_{ij} {\bf S}_i \cdot {\bf S}_j + K_{ij} \left( {\bf S}_i \cdot {\bf S}_j \right)^2 \right] \nonumber \\ 
 &-& \sum_{\triangpic}  \mathfrak{R}_3 \left( P_{123} + \Hc \right) + \sum_{\rhombpic} \mathfrak{R}_4  \left( P_{1234} + \Hc \right), 
\label{ham_ring}~~~~~
\end{eqnarray}
where ${\bf S}_{i}$ is the local moment at site $i$ of the Fe square lattice, and the bilinear and biquadratic interactions are chosen up to the third neighbors.
The $P_{123}$ and $P_{1234}$ stand for the $3$-site and $4$-site ring exchanges with $\mathfrak{R}_3, \mathfrak{R}_4 >0$ \cite{Lai_SU(3)_gapless}, which rotate the states such as $P_{123} | \alpha \beta \gamma\ra = |\gamma \alpha \beta\ra,~P_{1234} | \alpha \beta \gamma \xi \ra = | \xi \alpha \beta \gamma \ra$. 
The ring exchange operators can be re-expressed via physical spin operators \cite{Itoi97} as detailed in Supplemental Material \cite{supplemental}. The biquadratic terms also introduce the quadrupolar operator ${\bf Q}_i$, which has five components: $Q^{x^2 - y^2}_i = (S^x_i)^2 - (S^y_i)^2$, $Q^{3z^2 - r^2}_i = [ 2 (S^z_i)^2 - (S^x_i)^2 - (S^y_i)^2]/\sqrt{3}$, $Q^{xy} = S^x_i S^y_i + S^y_i S^x_i$, $Q^{yz} =  S^y_i,S^z_i +S^z_i S^y_i$, and, $Q^{zx} = S^z_i S^x_i + S^x_i S^z_i$. The biquadratic term can be re-expressed as $({\bf S}_i \cdot {\bf S}_j )^2 = ({\bf Q}_i \cdot {\bf Q}_j) /2 - ({\bf S}_i \cdot {\bf S}_j) /2 + ({\bf S}^2_i {\bf S}^2_j)/3$.

\textit{Semi-classical phase diagram for spin-$1$ model.}---
We first use the site-factorized wavefunction analysis to study the semi-classical phase diagram for $S = 1$. Based on the time-reversal invariant basis of the SU(3) fundamental representation, $
|x\ra = \left[ i |1\ra - i |\bar{1}\ra \right]/\sqrt{2}, ~ |y\ra = \left[ |1\ra + |\bar{1}\ra \right]/\sqrt{2},~~ |z\ra = -i |0\ra,$ where we abbreviate $|S^z = \pm1\ra \equiv |\pm1\ra$, $|S^z = 0\ra \equiv |0\ra$, and $|\bar{1}\ra \equiv |-1\ra$, we can introduce the complex site-factorized wavefunction vectors at each site to characterize any possible ordered state with short-ranged correlations as $ {\bf d}_i =  ( d^x_i, d^y_i, d^z_i  )$ in the $\{|x\ra, |y\ra, |z\ra\}$ basis (${\bf d}_i = {\bf u}_i + i {\bf v}_i$). The normalization of the wavefunction leads to
 the constraint ${\bf d}_i \cdot \bar{\bf d}_i = 1$, or equivalently, ${\bf u}^2_i + {\bf v}^2_i =1$, and the overall phase can be fixed by requiring ${\bf d}_i^2 = \bar{\bf d}_i^2$, i.e., ${\bf u}_i \cdot {\bf v}_i = 0$. In terms of ${\bf d}$, the Hamiltonian~\eqref{ham_ring} is expressed as
\begin{eqnarray}
&&H = \sum_{i, \delta_n} \left[J_n\left| {\bf d}_i \cdot \bar{\bf d}_j\right|^2
+  \left(K_n - J_n\right) \left|{\bf d}_i \cdot {\bf d}_j \right|^2 + K_n\right]~~~~ \nonumber \\ 
&& - \sum_{\triangpic}  \mathfrak{R}_3 (\bf{d}_1 \cdot \bar{\bf{d}}_3) ( \bf{d}_2 \cdot \bar{\bf{d}}_1) ( \bf{d}_3 \cdot \bar{\bf{d}}_2) + \Hc \nonumber \\ 
&&  + \sum_{\rhombpic} \mathfrak{R}_4 (\bf{d}_1 \cdot \bar{\bf{d}}_4) (\bf{d}_2 \cdot \bar{\bf{d}}_1)(\bf{d}_3 \cdot \bar{\bf{d}}_2) (\bf{d}_4 \cdot \bar{\bf{d}}_3) + \Hc. \label{Hd_4ring}~~~~~
\end{eqnarray}
In the following, we will drop the irrelevant constant terms.
\begin{figure}[t]
    \subfigure[]{\label{Fig:PD_noring}\includegraphics[width= 1.5 in]{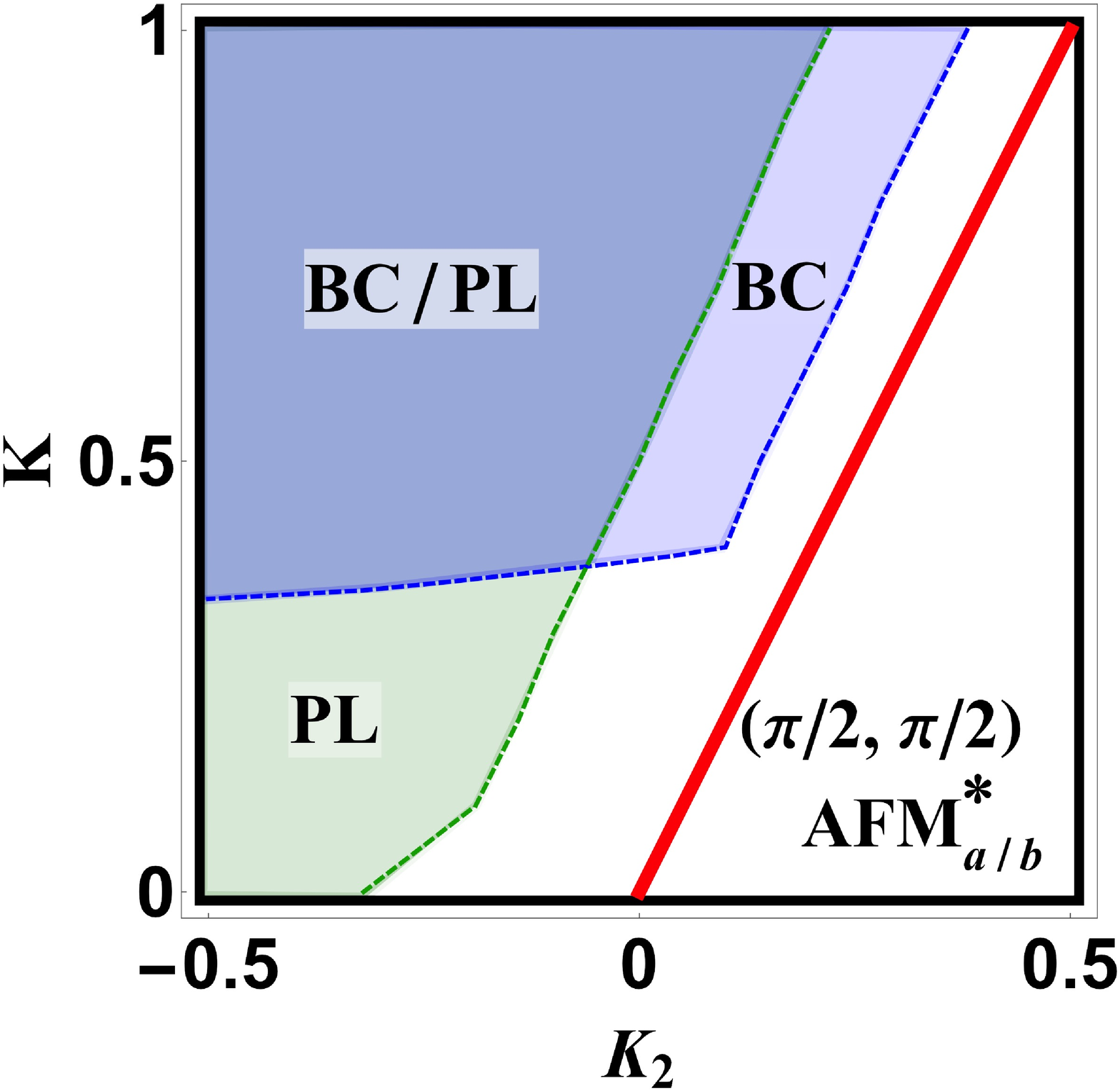}}
    \subfigure[]{\label{Fig:FLWT_lattice}\includegraphics[width=1.3 in]{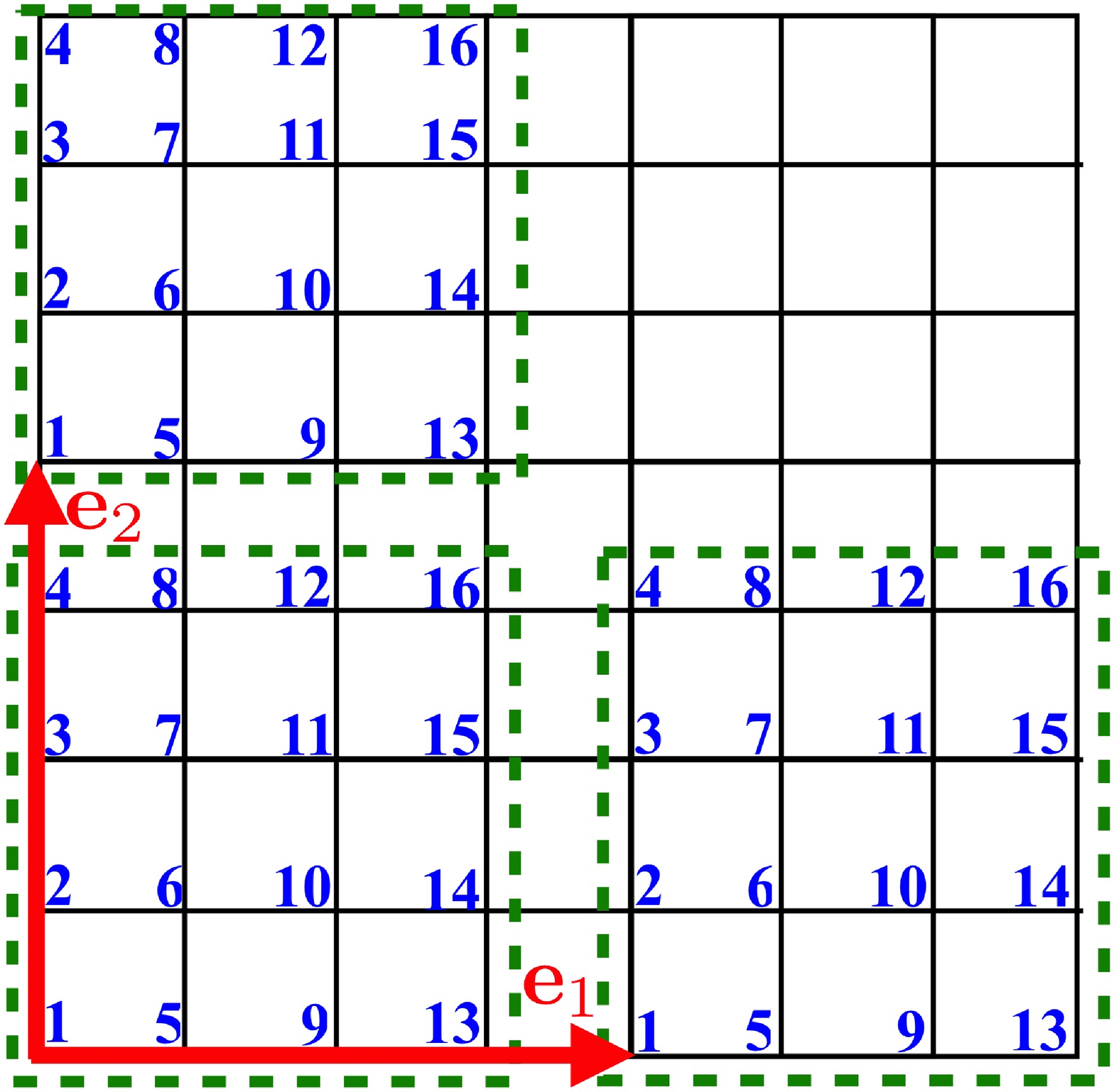}}
    \caption{(color online) (a) Site-factorized wavefunction phase diagram in the absence of ring exchange terms for spin-$1$. Here we fix $(J_1, J_2, J_3, \mathfrak{R}_3,\mathfrak{R}_4) = (1, 0.8,1, 0, 0)$ and vary $K_1 = K_3=-K$, $K_2 $. The red line represents the boundary between BC/PL and $(\pi/2, \pi/2)$ AFM$^*_{a/b}$ determined by the site-factorized energetics, $K= 2K_2$. The blue and green shaded regimes are the regimes of BC and PL with \textit{nonzero stiffness} determined by the spin-$1$ flavor-wave theory. (b) Illustration of the square network consisting of the lattice for performing flavor-wave theory calculation. There are $16$ sublattices and the unit cells are connected by the vectors ${\bf e}_1 \equiv \hat{x}$ and ${\bf e}_2 \equiv \hat{y}$. }
\label{Fig:PD_FLWT}
\end{figure}

To illustrate the energetic degeneracy of the BC and PL orders, we first set ring exchanges $\mathfrak{R}_3, \mathfrak{R}_4=0$ and without loss of generality we study the $J$-$K$ model with $(J_1,J_2, J_3) = (1, 0.8, 1)$. We vary $K_1 = K_3\equiv -K <0$ and $K_2 \in [-0.5,0.5]$ on an $L\times L$ square lattice with $L$ up to $8$ to obtain the site-factorized wavefunction phase diagram, Fig.~\ref{Fig:PD_noring}. Besides the intriguing doubly degenerate BC/PL orders, we find another doubly degenerate orders dubbed $(\pi/2, \pi/2)$ AFM$^*_{a/b}$ and abbreviated as AFM$^*_{a/b}$ below, whose spin patterns  are illustrated in Figs.~\ref{Fig:pi/2pi/2_phase}(c)-(d). \cite{AFMs_note} The semi-classical energies for these two orders in the absence of the ring exchange interactions are $ \mathcal{E}_{BC/PL}= K_1  + K_2 + 2(K_3 - J_3)$ and $ \mathcal{E}_{AFM_{a/b}^*} = \frac{3}{4}K_1 + \frac{1}{2} K_2 + 2(K_3 - J_3)$. This gives the phase boundary, $K_1 + 2K_2 =0$, consistent with the numerical result in Fig.~\ref{Fig:PD_noring}.

The BC/PL orders have different ``spin  dipolar nematicity" along the off-diagonal directions of the square lattice, defined as $\sigma^s_{2} \equiv (1/N_s) \sum_j {\bf S}_j \cdot \left[{\bf S}_{j + \hat{x} + \hat{y}} -  {\bf S}_{j - \hat{x} + \hat{y}}\right]$ (where $N_s$ is the total number of lattice site). This nematic order parameter is nonzero for the BC phase, but vanishes for the PL phase. The two orders, AFM$^*_{a/b}$, do not possess the spin dipolar nematicity. Still, they have broken $C_4$ symmetry due to the nonzero ``spin quadrupolar nematicity", defined in terms of $\sigma^Q_1 = 1/N_s \sum_j {\bf Q}_j \cdot \left[ {\bf Q}_{j + \hat{x}} - {\bf Q}_{j + \hat{y}}\right]$. The spin quadrupolar nematicity along off-diagonal direction are also nonzero but are much smaller. 

Since our main focus is to examine the mechanism for stabilizing the BC order relevant to FeTe, below we focus on the BC/PL regime.

\textit{DMRG calculations for the $J$-$K$ model.}---
To analyze the robustness of the degeneracy in the BC and PL orders, we next turn to analyzing the quantum fluctuations in an unbiased way using the DMRG \cite{White1992} method with spin rotational SU(2) symmetry \cite{Mcchlloch2002, gong2014square}. We study cylinder system on two different geometries--the rectangular cylinder (RC) and the $\pi/4$-rotated tilt cylinder (TC) \cite{DMRG}. We show the spin correlations for spin-$1$ and spin-$2$ models at the parameter $(J_1,J_2,J_3,K_1,K_2,K_3)=(1,0.8,1,-0.5,0.1,-0.5)$ in Fig.~\ref{Fig:DMRG_main} (see Supplemental Material for spin-$3/2$ \cite{supplemental}). 

For the models with different spin-$S$, we find the PL order on the RC cylinder and the BC order on the TC cylinder. On a finite-size system, different states being stabilized on different geometries suggest energetically degenerate competing states \cite{gong2015honeycomb}. This is corroborated by a comparison of the bulk energies of the two states on different system sizes \cite{supplemental}. On the large cylinders, we find the energies of the two states to be quite close to each other. For the spin-$1$ model, the energy difference between the RC ($L_y = 8$) and TC ($L_y = 6$) cylinders is only $0.2\%$. Thus, our DMRG results strongly suggest the (quasi-) degeneracy of the BC and PL states  in the $J$-$K$ model, even after the quantum fluctuations effects are considered, which is also consistent with the flavor-wave theory results to be presented below.
\begin{figure}[t]
    \subfigure[]{\label{Fig:L8a}\includegraphics[width= 1 in]{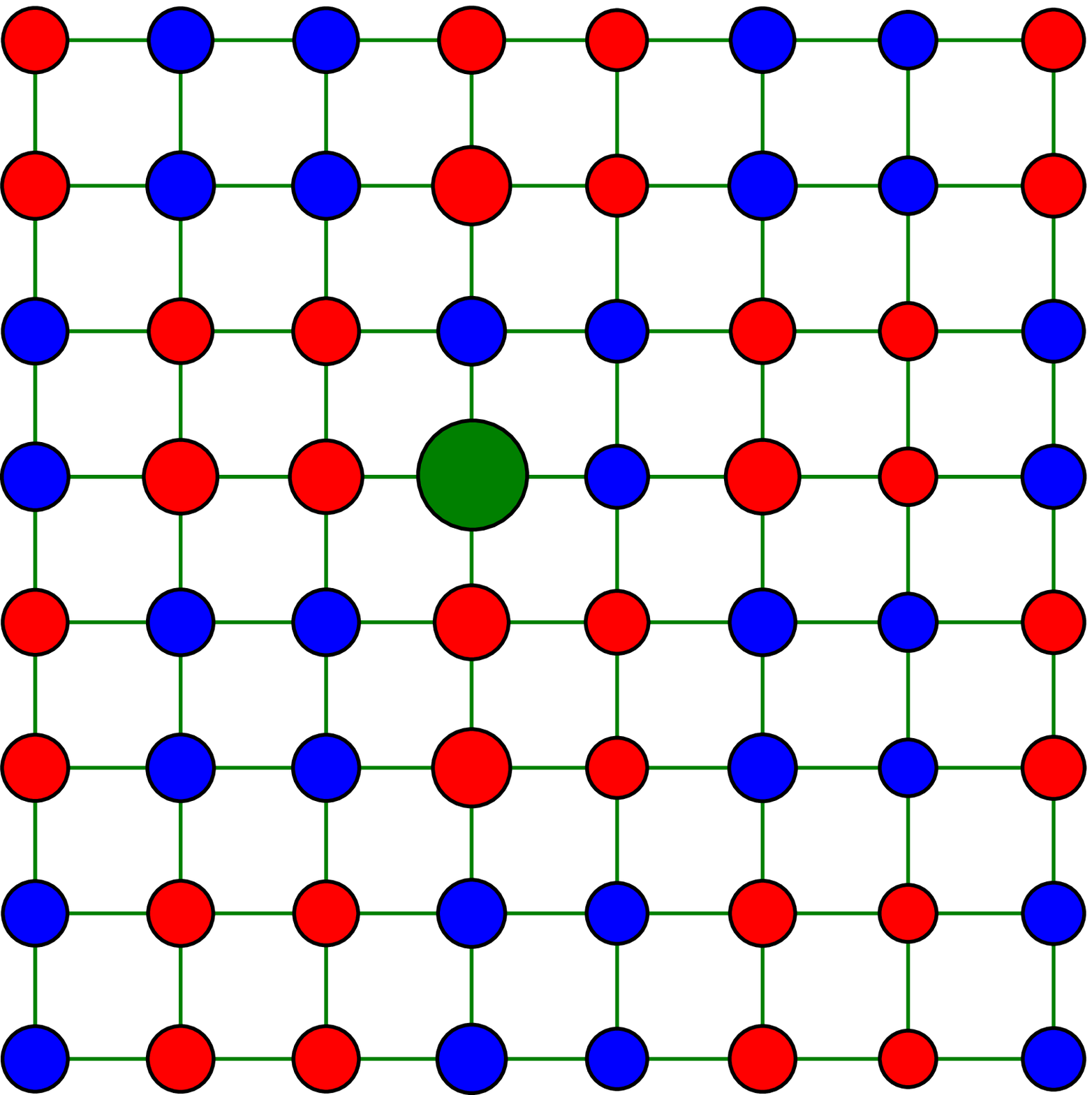}}~~~~
    \subfigure[]{\label{Fig:L8b}\includegraphics[width=1 in]{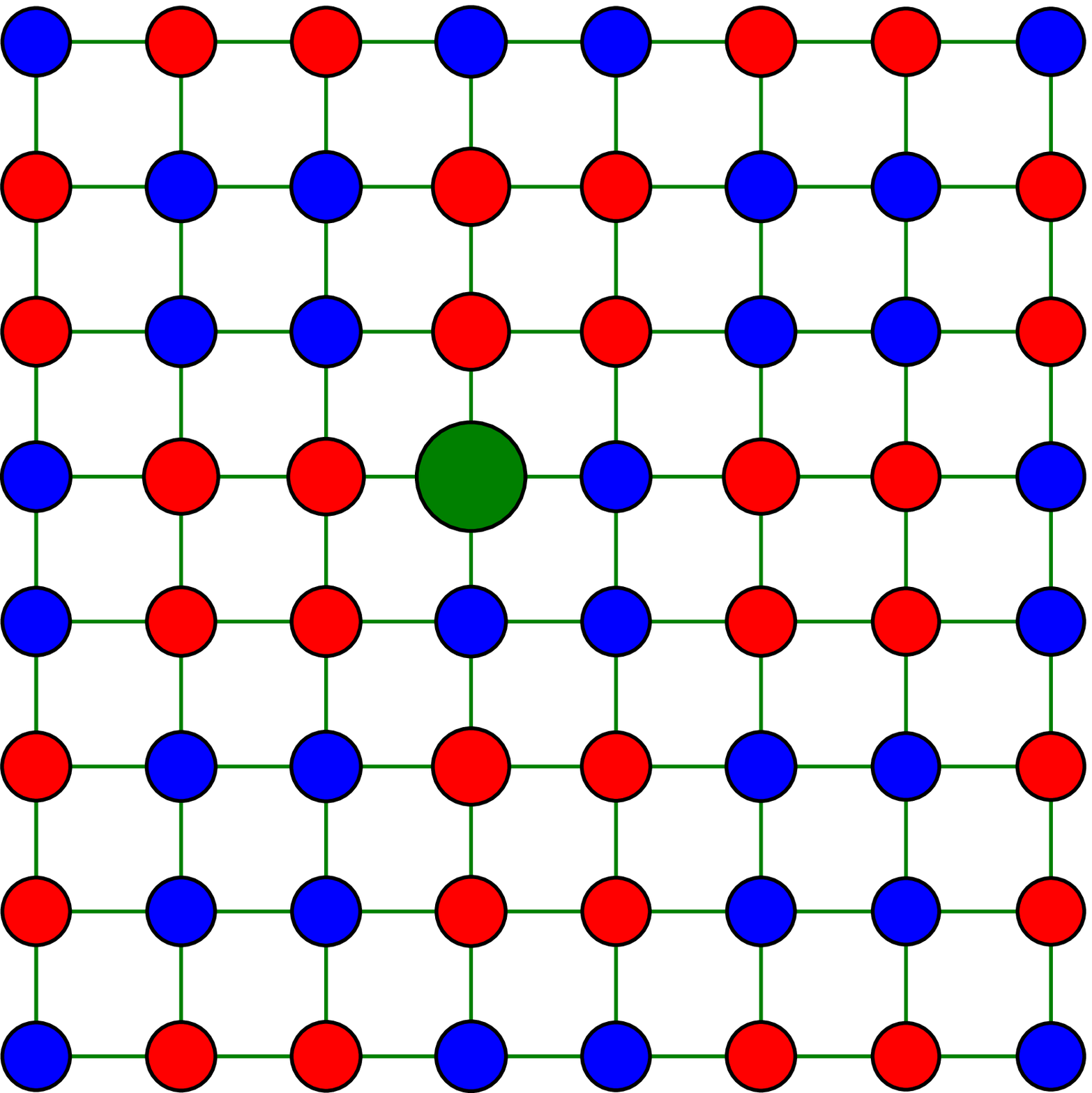}}\\
     \subfigure[]{\label{Fig:L8c}\includegraphics[width= 1 in]{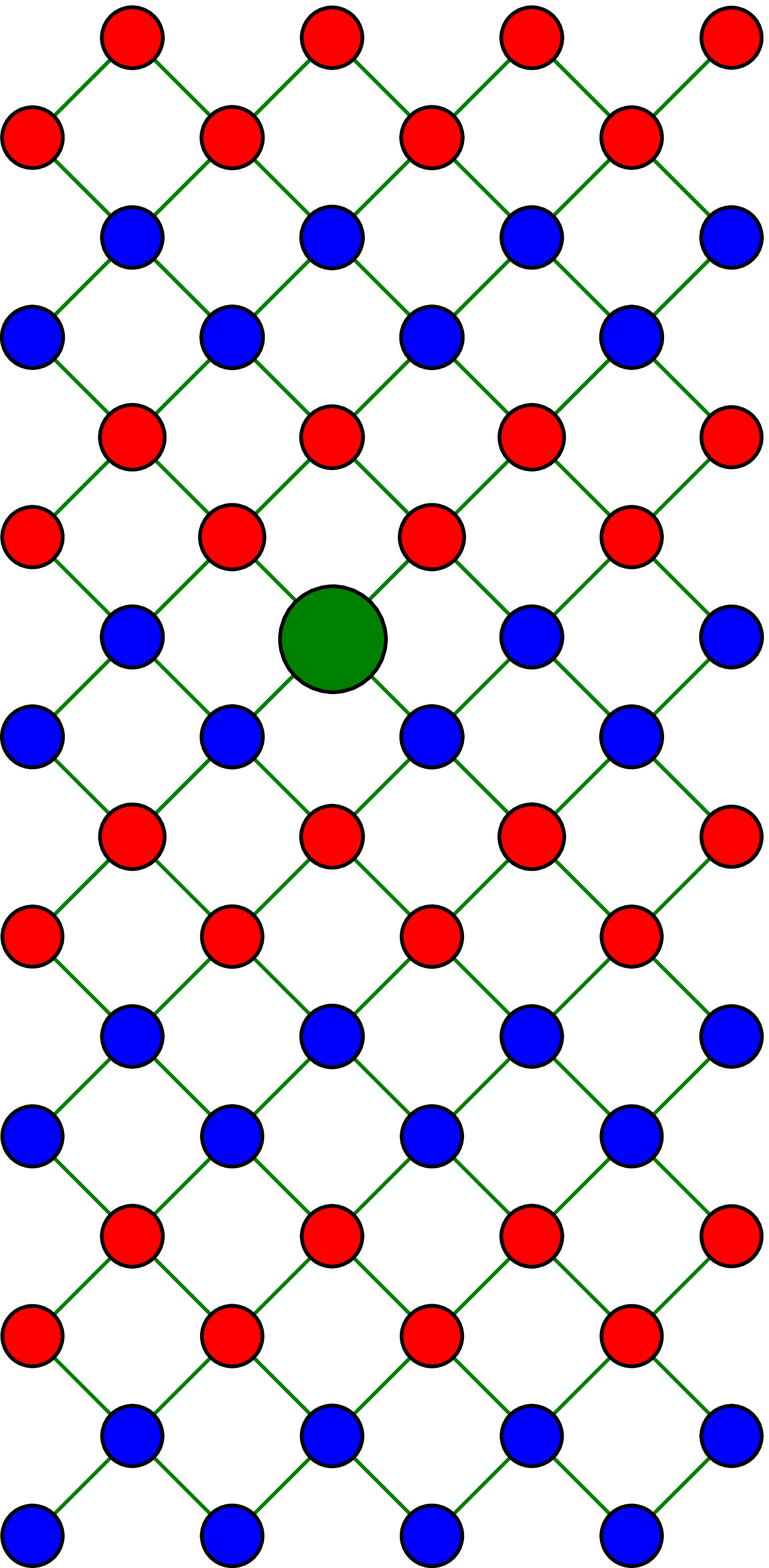}}~~~~
    \subfigure[]{\label{Fig:L8d}\includegraphics[width=1 in]{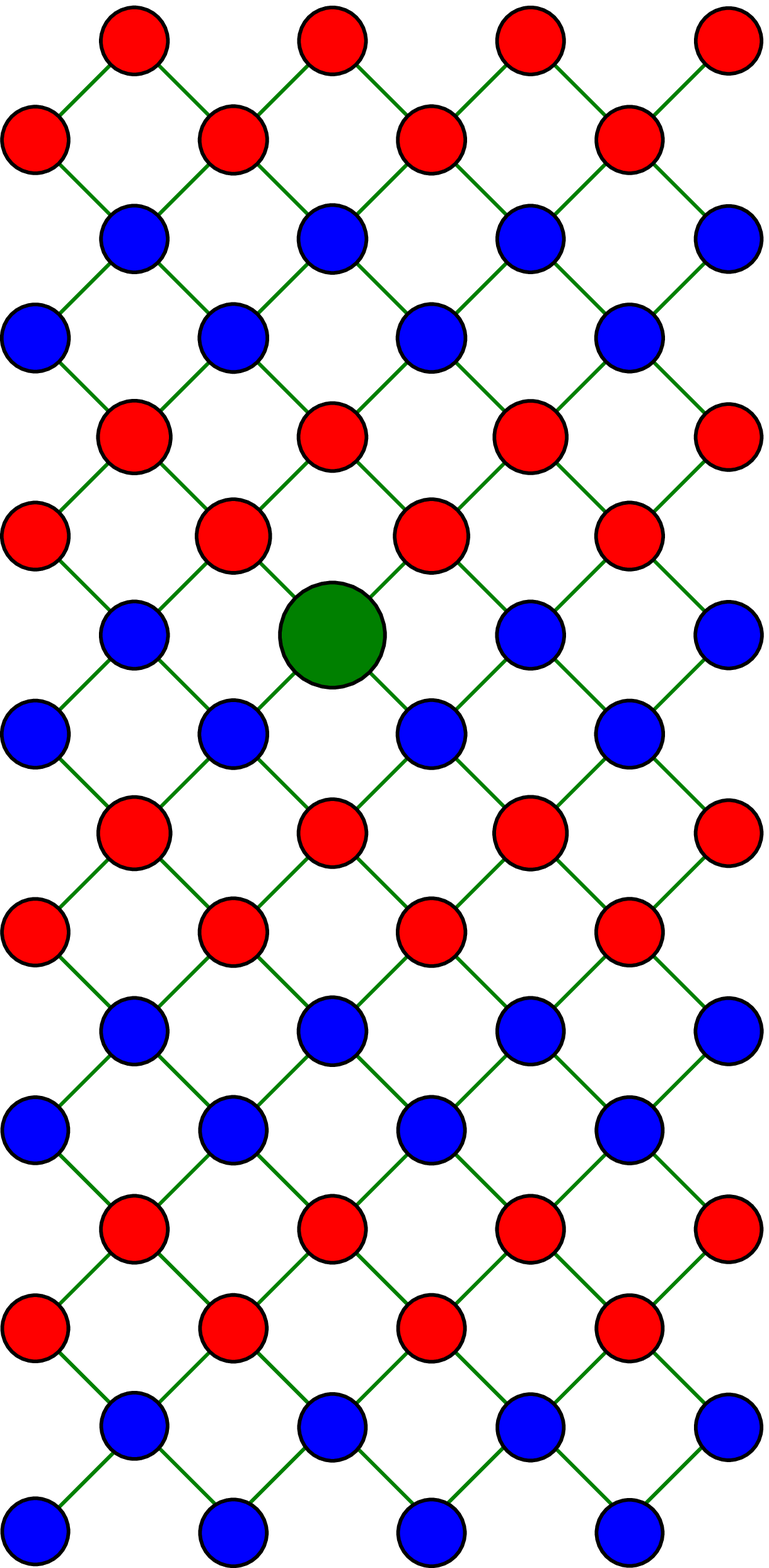}}
    \caption{(color online) Spin correlation in the middle of cylinders $(L_x=24, L_y=8)$ for the $(\pi/2,\pi/2)$ phase at $J_1=1.0$, $J_2=0.8$, $J_3=1.0$, $K_1=K_3=-0.5$, and $K_2=0.1$ on rectangular cylinder (RC) and tilted cylinder (TC). (a) and (c) are for $S=1$; (b) and (d) are for $S=2$. The green site is the reference site; the blue and red colors denote positive and negative correlations of the sites with the reference site, respectively. The circle radius is proportional to the magnitude of spin correlation.}
\label{Fig:DMRG_main}
\end{figure}

\textit{Stiffness of the BC/PL orders in the $J$-$K$ model.}---
To address the mechanisms for breaking the BC/PL degeneracy in the BC and PL phases of the $J$-$K$ model, we perform the flavor-wave theory analysis \cite{Bauer2012, Lai_SU(3)_gapless}, which considers partial quantum fluctuations going beyond the semi-classical picture. Within the flavor wave calculations for BC and PL, we consider a square lattice consisting of $16$ sublattice per unit cell illustrated in Fig.~\ref{Fig:FLWT_lattice}. In addition, we associate $3$ Schwinger-bosons at each site $i$, $b_{i\alpha=x,y,z}$, to the states under SU(3) time-reversal invariant basis, $|x\ra,~|y\ra,~|z\ra$, where $b^\dagger_{i\alpha} |vac\ra = |\alpha\ra$ with $|vac\ra$ being the vacuum state of the Schwinger bosons. The bosons satisfy a local constraint, $\sum_{\alpha} b^\dagger_{i \alpha}b_{i \alpha} = 1.$ The model Hamiltonian in terms of spin operators can be re-expressed in terms of the bosons,
\begin{eqnarray}
H \simeq \sum_{i, \delta_n, \alpha, \beta} \left[ J_nb^\dagger_{i \alpha} b_{i \alpha} b^\dagger_{j\beta} b_{i\beta} + \left(K_n - J_n\right) b_{i\alpha}^\dagger b^\dagger_{j \alpha} b_{i\beta} b_{i\beta}\right],~~~
\end{eqnarray}
where $j = i + \delta_n,$ and $\delta_n$ (with $n = 1,~2,~3$) connects site $i$ to its $n$th nearest neighbor sites, and we have ignored the constant terms $K_n - J_n$ in the above equation. We also introduce different local rotations for different site $j$. For site $j \in | S^z = \pm1\ra$, we introduce
\begin{eqnarray}
\begin{pmatrix}
b_{jx} \\
b_{jy} \\
b_{jz}
\end{pmatrix} =
\begin{pmatrix}
\mp \frac{i}{\sqrt{2}} & \frac{1}{\sqrt{2} }& 0\\
\frac{1}{\sqrt{2}} & \mp \frac{i}{\sqrt{2}} & 0 \\
0 & 0 & 1
\end{pmatrix}
\begin{pmatrix}
d_{jx}\\
d_{jy}\\
d_{jz}
\end{pmatrix},
\end{eqnarray}
where for each site there is only one flavor of bosons $d_{ix}$ condensing, 
and we replace $d^\dagger_{ix}$ and $d_{ix}$ by $(M - d_{iy}^\dagger d_{iy} - d^\dagger_{iz} d_{iz})^{1/2}$, 
with $M=1$ in the present case. A $1/M$ expansion up to the quadratic order of the Holstein-Primakoff bosons 
$d_y$ and $d_z$ followed by a Bogoliuobov transformation allows us to extract the ground state energy \cite{Xiao2009}.
The detailed derivations are presented in the Supplemental Material \cite{supplemental}. Numerically diagonalizing the quadratic boson Hamiltonians on a square network consisting of $100\times 100$ cluster unit cells gives the energies of BC and PL which can be used to extract the regimes of BC and PL with finite stiffness. 

The results are summarized in the shaded regimes in Fig.~\ref{Fig:PD_noring}. 
The blue/green shaded regions in Fig.~\ref{Fig:PD_noring} represent the stable BC/PL regimes in which the stiffness of BC/PL are finite. Surprisingly, we find that the BC and PL have \textit{different} stiffness. In particular, there is a parameter range over which the BC order is stable, while the PL order is not due to negative stiffness (see supplementa material \cite{supplemental}). This represents a fluctuation mechanism to select the BC AFM order.

We find a much wider parameter regime where both BC and PL have positive stiffness. The flavor-wave theory also suggests that the energy splitting between these two orders are negligible (for a point deep inside this regime, the energy splitting is $\leq 0.1\%$; near the boundary, the energy splitting is $\sim 1\%$). Because this common regime, where both orders are stable, covers a large parameter space, we need to explore additional inputs 
for another mechanism that  breaks the degeneracy and stabilizes the BC. This leads us to discuss the effect 
of ring exchanges in the next subsection.

We close this subsection with a remark on the limit of vanishing $K,~K_2$ (and in the absence of the ring exchange interactions); further details are given in the Supplemental Material \cite{supplemental}. For $(J_1, J_2, J_3) = (1, 0.8, 1)$, which we have so far focused on, Fig.~\ref{Fig:PD_noring} shows that the BC/PL phases are unstable at $K, K_2 \rightarrow 0$. In this limit, the flavor-wave theory suggests that neither of the BC/PL phases can be stable except for infinitesimally small ratios of $J_1/J_3$ and $J_1/J_2$ in the $J_1$-$J_2$-$J_3$-only model. Consider, for example, $(J_2, J_3) =( 0.8,1.0)$, $K,K_2 = 0$: we find that the BC and PL can only be stable for a very small range of  $J_1 \leq 0.03$.
(Within this regime, the energies between the two orders are degenerate, with their difference being $\sim 0.001\%$.)
This implies that the previous conclusion suggesting that $1/S$ quantum fluctuations 
stabilize the PL phase in an extended parameter range cannot apply to the $J_1$-$J_2$-$J_3$ model \cite{Chubukove_PL2012}. Going ``slightly'' beyond the $J$-only model by considering the effects of $K_1 = K_3 = -K$ with $K_2=0$, we find that increasing $K$ significantly enhances the threshold values of $J_1$ for stable BC and PL orders. Taking $K=0.1, K_2 =0$, we find threshold values $J^{BC}_1 \simeq 0.42$ for BC and $J^{PL}_1 \simeq 0.5$ for PL (see Supplemental Material \cite{supplemental}).
 On the other hand, increasing $K_2 > 0$ reduces the stable BC and PL regimes. 
 
\textit{Ring exchanges stabilizing BC.}---
We are now in position to address the role of the ring exchange interactions in 
the wide parameter regime where the BC and PL phases are degenerate, with both having positive stiffness. In the presence of the ring exchanges, the energy corrections, within the site-factorized wavefunction, to each order are
\begin{eqnarray}
 &\mathcal{E}^{BC}_\mathfrak{R} =  -2 \mathfrak{R}_3,~~~&\mathcal{E}^{PL}_\mathfrak{R} = -2 \mathfrak{R}_3 + \frac{\mathfrak{R}_4}{2},\\
&\mathcal{E}^{AFM^{*}_a}_\mathfrak{R} = - \mathfrak{R}_3, ~~~& \mathcal{E}^{AFM^{*}_b}_\mathfrak{R} = - \mathfrak{R}_3 + \frac{\mathfrak{R}_4}{4}.
\end{eqnarray} 
The $3$-site ring exchange interaction, $\mathfrak{R}_3$,
does not split the degeneracy of BC/PL and that of AFM$^*_{a/b}$, although
it does lower the energies of BC/PL and AFM$^*_{a/b}$ by different amounts and make those of BC/PL lower. 
By contrast, the $4$-site ring exchange, $\mathfrak{R}_4$,
 \textit{does} lift the degeneracy between BC/PL (as well as that between AFM$^*_{a/b}$).

\begin{figure}[t] 
   \centering
   \includegraphics[width=2 in]{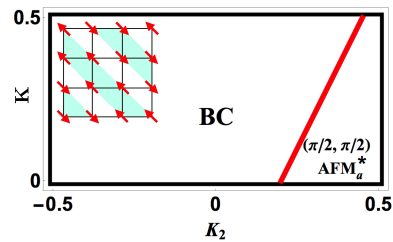} 
   \caption{(color online) The site-factorized wavefunction phase diagram with small ring exchange terms. Here we fix $(J_1, J_2, J_3, \mathfrak{R}_3,\mathfrak{R}_4) = (1, 0.8,1, 0.1, 0.1)$ and vary $K_1 = K_3=-K$ and $K_2$. The $4$-site ring exchange interaction is responsible for lifting the degeneracy between BC/PL and that betweenAFM$^*_{a/b}$.}
   \label{Fig:PD_wring}
\end{figure}

To illustrate such effects of the ring exchanges, we take relatively small values $\mathfrak{R}_3 = \mathfrak{R}_4 = 0.1 \ll J_n, | K_n |$. The phase diagram of the $J$-$K$-$\mathfrak{R}$ model for this case is shown in Fig.~\ref{Fig:PD_wring}. The red line is the phase boundary with $K - 2K_2 = -0.4$. We note that, in Fig.~\ref{Fig:PD_wring}, the BC order will remain stable over an extended regime in the presence of quantum fluctuations, based on what we have established in Fig.~\ref{Fig:PD_noring}. Moreover, the energy splitting between BC and PL due to the presence of ring exchanges can already be observed at semi-classical level and we expect the splitting will be further enhanced if additional quantum fluctuation effects are incorporated by, \textit{e.g.}, the DMRG method.

\textit{Discussions.}---
We close by remarking on several points. The small ring exchanges do not 
affect dramatically the energetics of $(\pi,0)$ collinear AFM phase and the $(\pi,0)$ antiferro-quadrupolar order (AFQ) that are suggested to play major roles in the normal state of the iron chalcogenide FeSe \cite{YuSi_AFQ, Lai2016, Wenjun_nematic2016}. Indeed, at the semi-classical level, there are no
corrections to the energies of these states from the ring exchanges.

From the above, we can conclude that the frustrated magnetism encoded in the Hamiltonian, Eq.~\eqref{ham_ring}, is sufficiently rich to understand the magnetism and nematic properties of FeTe on the same footing with
those of FeSe and the iron pnictides. Given the strong indication for the importance of the spin physics to the iron-based superconductivity \cite{PCDai2015, Si2016},
the unification we have achieved here represents an important virtue of the mechanism we have advanced in the present work.

Finally, it is worth noting that alternative mechanisms for the BC in FeTe have assumed that the spin physics itself does not yield the ($\pi/2$,$\pi/2$) bicollinear antiferromagnetic state, but such an order arises under additional FeTe-specific driving forces beyond the spin physics. The additional driving forces that have been discussed include an orbital order \cite{Turner2009, Singh2009} or spin-lattice interactions \cite{Fang2009,EDagotto_FeTe}.

\textit{Conclusion.}---
We have studied the quantum bilinear-biquadratic model in the presence of ring exchange interactions, and identified two mechanisms that stabilize the $(\pi/2,\pi/2)$ bicollinear antiferromagnetic order experimentally observed in FeTe. In the absence of ring exchanges, we demonstrate a relatively narrow parameter range where quantum fluctuations select the bicollinear order, by destabilizing the classically-degenerate $(\pi/2,\pi/2)$ plaquette antiferomagnetic order. 
We also identify a larger parameter range where the bicollinear and plquettes orders are degenerate and both are stable. In this regime, the presence of a $4$-site ring exchange interaction breaks the degeneracy and selects the bicollinear order. Because the second mechanism operates over a considerably more extended parameter regime, it represents a more robust 
mechanism to understand the magnetic and nematic orders of FeTe. Our work unifies the electronic order of FeTe with those of the other iron chalcogenides and iron pnictides within a single Hamiltonian, and highlights the importance of spin frustration to the magnetism and superconductivity of the iron-based systems.

\textit{Acknowledgement.}---
The work was supported in part by the NSF Grant No.\ DMR-1611392 and the Robert A.\ Welch Foundation Grant No.\ C-1411 (W.-J.H., H.-H.L. and Q.S.), the NSF Grant No.\ DMR-1350237 (W.-J.H. and H.-H.L.), a Smalley Postdoctoral Fellowship of the Rice Center for Quantum Materials (H-H. L.), the National High Magnetic Field Laboratory through the NSF Grant No.\ DMR-1157490 and the State of Florida (S.-S.G.). The majority of the computational calculations have been performed on the Shared University Grid at Rice funded by NSF under Grant EIA-0216467, a partnership between Rice University, Sun Microsystems, and Sigma Solutions, Inc., the Big-Data Private-Cloud Research Cyberinfrastructure MRI-award funded by NSF under Grant No. CNS-1338099 and by Rice University, the Extreme Science and Engineering Discovery Environment (XSEDE) by NSF under Grants No.\ DMR160003 and DMR160057. Computational support has also been provided by XSEDE from the NSF under Grant No.\ DMR160004 (S.-S.G.).

\bibliography{bib4FeTeBC}

\begin{thebibliography}{42}
\bibitem [{AFM()}]{AFMs_note}%
  \BibitemOpen
  \href@noop {} {}\bibinfo {note} {{The directions of the nearest-neighbor
  spins can be tilted by an angle $\theta$ in a way that all second-neighbor
  spins remain orthogonal, while the third-neighbor spins are always parallel.
  Despite of $\theta$, the energy of the AFM$_{a/b^*}$ is independent of
  $\theta$.}}\BibitemShut {Stop}%
\end{thebibliography}

\clearpage
\begin{widetext}
\section{Supplemental Material}
\section{Ring exchanges and spin operators} \label{App:ring}
The $3$-site and $4$-site ring exchanges $P_{123}$ and $P_{1234}$ are equivalently expressed in terms of the physical spin operators. In general ground, the ring exchange $P_{ijk\cdots \ell}$ can be expressed in terms of the products of two-site exchange operator $P_{ijk\cdots \ell m} = P_{ij} P_{jk} \cdots P_{\ell m}$. The two-site ring exchange operator can be written in terms of physical spin operators. Defining $X\equiv {\bf S}_i \cdot {\bf S}_j$, we can specify the results as follows \cite{Itoi97}
\begin{eqnarray}
&S=\frac{1}{2}: ~~~& ~~~ P^{1/2}_{ij} = 2X + \frac{1}{2},\\
&S=1:~~~& ~~~ P^1_{ij} = X^2 + X -1,\\
&S=\frac{3}{2}:~~~& ~~~ P^{3/2}_{ij} = \frac{2}{9} X^3 + \frac{11}{18}X^2 - \frac{9}{8}X - \frac{67}{32},\\
&S = 2:~~~& ~~~ P^2_{ij} = \frac{1}{36} X^4 + \frac{1}{6} X^3 - \frac{7}{12}X^2 - \frac{5}{2} X -1.
\end{eqnarray}

For studying the $S=1$ system in the current work, we introduce a site-factorized wavefunction in the time-reversal-invariant SU(3) basis, $\{ |x\ra, |y\ra, |z\ra \}$ as a complex vector ${\bf d}_j = {\bf u}_j + i {\bf v}_j$ at site $j$. Within the site-factorized wavefunction analysis, the wavefunction of $n$-sites is simply the tensor product of the wavefunction at each site. The ring exchange operator cyclically rotates the states between these sites. 
\begin{eqnarray}
&& P_{12\cdots n} |\Psi\ra_{12\cdots n} = P_{12\cdots n} \left[ {\bf d}_1 \otimes {\bf d}_2\otimes \cdots \otimes {\bf d}_n \right] = {\bf d}_n \otimes {\bf d}_1 \otimes {\bf d}_2 \otimes \cdots {\bf d}_{n-1}.
\end{eqnarray}
Therefore, within the site-factorized wavefunction, the $n$-site ring exchange terms contribute to the Hamiltonian as
\begin{eqnarray}
&& _{12\cdots n}\la \Psi| P_{12\cdots n} | \Psi \ra_{12\cdots n} = \left[ \bar{\bf d}_1 \otimes \bar{\bf d}_2 \otimes \cdots \otimes \bar{\bf d}_n \right] \cdot \left[   {\bf d}_n \otimes {\bf d}_1 \otimes \cdots \otimes \bar{\bf d}_{n-1} \right] = (\bar{\bf d}_1 \cdot {\bf d}_n )(\bar{\bf d}_2 \cdot {\bf d}_1) \cdots (\bar{\bf d}_n \cdot {\bf d}_{n-1}) .
\end{eqnarray}
The cases of $n=3,4$ correspond to those shown in the main text.

\section{Site-factorized wavefunction energies of competing orders}\label{App:SFW}
In the present work, within the site-factorized wavefunction studies we find four competing phases, $(\pi/2,\pi/2)$ bicollinear order(BC), $(\pi/2,\pi/2)$ plaquette order(PL), $(\pi/2,\pi/2)$ AFM$_a^*$, and $(\pi/2,\pi/2)$ AFM$_b^*$, in the $J$-$K$-$\mathfrak{R}$ model as
\begin{eqnarray}
&& \mathcal{E}_{BC}= K_1  + K_2 + 2(K_3 - J_3)- 2\mathfrak{R}_3,\\
&& \mathcal{E}_{PL}= K_1  + K_2 + 2(K_3 - J_3) - 2\mathfrak{R}_3+ \frac{1}{2}\mathfrak{R}_4,\\
&& \mathcal{E}_{(\pi/2, \pi/2)AFM_a^*} = \frac{3}{4}K_1 + \frac{1}{2} K_2 + 2(K_3 - J_3) -\mathfrak{R}_3, \\
 && \mathcal{E}_{(\pi/2, \pi/2)AFM_b^*} = \frac{3}{4}K_1 + \frac{1}{2} K_2 + 2(K_3 - J_3) -\mathfrak{R}_3+ \frac{1}{4}\mathfrak{R}_4.
\end{eqnarray}
In the absence of the ring exchanges, $\mathfrak{R}_{3/4}$, BC and PL are degenerate and so are $(\pi/2,\pi/2)$ AFM$_a^*$ and $(\pi/2,\pi/2)$ AFM$^*_b$. The boundary between degenerate BC/PL and degenerate $(\pi/2,\pi/2)$ AFM$_{a/b}^*$ in the absence of ring exchange terms is determined by $K_1 = -2 K_2 \Rightarrow K = 2 K_2$, where we explicitly use the definition $K_1  = -K$. We remark that the site-factorized wavefunction analysis in the present spin-$1$ model suggests that in the regime of $(\pi/2,\pi/2)$ AFM$^*_{a/b}$ the nearest-neighbor spins be tilted on the plane from the spin patterns shown in Figs. 1(c)-(d) by a small angle $\theta$ in such a way that each second-neighbor spins remain orthogonal, while third-neighbor spins are always anti-parallel to each other. Despite of the angle $\theta$, the energy of AFM$_{a/b}^*$ is independent of $\theta$. 

We note that in numerics, there are indeed some fluctuations in the quadrupolar degrees of freedom and the dipolar magnetization can be slightly smaller than one, $|\la{\bf S}\ra| \leq 1$. The energy difference between the phases obtained in the exact numerical result and the AFM$^*_{a/b}$ are within $O(10^{-2})$, which may be due to the numerical errors in searching for a global minimum in the $4\times L^2$ parameter space. On other other hand, this may suggests there are many competing local minima in the AFM$^*_{a/b}$ regime in the parameter space in the $S=1$ model. Enlarging the spin size $S >1$ sufficiently suppresses the fluctuations in the quadrupolar degrees of freedom and the spin pattern can be unbiasedly confirmed in the density matrix renormalization group analysis. \cite{FeTe-long} 

In the presence of ring exchanges, the $3$-site ring exchanges within the site-factorized wavefunction analysis can not split the degeneracies, but they lower the energies of BC/PL more than those of $(\pi/2,\pi/2)$ AFM$_{a/b}^*$. The $4$-site ring exchanges split the double degeneracy in both BC/PL and $(\pi/2,\pi/2)$ AFM$^*_{a/b}$. The competing phases become BC and $(\pi/2,\pi/2)$ AFM$_a^*$. The boundary can be determined by the equation, $K_1 + 2K_2 = 4 \mathfrak{R}_3 \Rightarrow K = 2K_2 - 4 \mathfrak{R}_3.$

\section{Flavor wave theory for $(\pi/2,\pi/2)$ bicollinear order and $(\pi/2,\pi/2)$ plaquette order}\label{App:FLWT}
Within the flavor wave calculation for BC and PL with $16$ sublattice per unit cell illustrated in Fig. 2(b) in the main text, we associate $3$ Schwinger-bosons at each site $i$, $b_{i\alpha=x,y,z}$, to the states under SU(3) time-reversal invariant basis, $|x\ra,~|y\ra,~|z\ra$, where $b^\dagger_{i\alpha} |vac\ra = |\alpha\ra$ with $|vac\ra$ being the vacuum state of the Schwinger bosons. The bosons satisfy a local constraint
\begin{eqnarray}\label{Eq:constraint}
\sum_{\alpha} b^\dagger_{i \alpha}b_{i \alpha} = 1.
\end{eqnarray}
The model Hamiltonian in terms of spin operators can be re-expressed in terms of the bosons,
\begin{eqnarray}
H = \sum_{i, \delta_n} \left[ J_n {\bf S}_i \cdot {\bf S}_j + K_n \left( {\bf S}_i \cdot {\bf S}_j \right)^2 \right] = \sum_{i, \delta_n, \alpha, \beta} \left[ J_nb^\dagger_{i \alpha} b_{i \alpha} b^\dagger_{j\beta} b_{i\beta} + \left(K_n - J_n\right) b_{i\alpha}^\dagger b^\dagger_{j \alpha} b_{i\beta} b_{i\beta}\right],
\end{eqnarray}
where $j = i + \delta_n,$ and $\delta_n$ (with $n = 1,~2,~3$) connects site $i$ to its $n$th nearest neighbor sites, and we ignore the constant terms $K_n - J_n$ in the above equation. For performing flavor wave theory calculation, we introduce different local rotations for different site $j$. For site $j \in | S^z = \pm1\ra$, we introduce
\begin{eqnarray}
\begin{pmatrix}
b_{jx} \\
b_{jy} \\
b_{jz}
\end{pmatrix} =
\begin{pmatrix}
\mp \frac{i}{\sqrt{2}} & \frac{1}{\sqrt{2} }& 0\\
\frac{1}{\sqrt{2}} & \mp \frac{i}{\sqrt{2}} & 0 \\
0 & 0 & 1
\end{pmatrix}
\begin{pmatrix}
d_{jx}\\
d_{jy}\\
d_{jz}
\end{pmatrix},
\end{eqnarray}
which still preserves the local constraint of Eq.~(\ref{Eq:constraint}) with $b_{i\alpha} \rightarrow d_{i\alpha}$. At each site only one flavor of bosons $d_{ix}$ condenses, and we replace $d^\dagger_{ix}$ and $d_{ix}$ by $(M - d_{iy}^\dagger d_{iy} - d^\dagger_{iz} d_{iz})^{1/2}$, with $M=1$ in the present case. A $1/M$ expansion up to the quadratic order of the Holstein-Primakoff bosons $d_y$ and $d_z$ followed by an appropriate transformation allows us to extract the ground state energy. We will replace the labeling $d_{i\alpha} = d_{\alpha}({\bf r},a)$, where ${\bf r}$ runs over the Bravais lattice of unit cells of the square network and $a=1,2,3,4,\cdots, 15,16$ runs over the sub lattices. The different unit cells are connected by ${\bf e}_1 \equiv \hat{x}$ and ${\bf e}_2 \equiv \hat{y}$.
 
For clarity, we introduce $ D^T_{\alpha = y,z} ({\bf k}) = \{d_\alpha({\bf k},1), d_\alpha({\bf k},2),\cdots,d_\alpha({\bf k},15),d_\alpha({\bf k},16)\}$, and $A_\alpha({\bf k}) = \left\{ D^T_\alpha({\bf k}), D^\dagger_\alpha(-{\bf k})\right\}$ for clarity below. Within the linear flavor wave theory calculation, we find that the Hamiltonian is $H_\mu = H_c + H^\mu_B$, where $\mu =$ BC or PL, and $ H_c = 32\sum_{\bf k} \left[ J_1 + J_2 - \frac{13}{4} J_3 + K_1 + K_2 + 3 K_3 \right]$, and can be determined straightforwardly independent of the boson fields and independent of BC or PL order. The $H^\mu_B \equiv \sum_{n=1,2,3}H^\mu_n$ represent the boson Hamiltonian related to the $n$th neighbor couplings that are different for $\mu=BC,PL$. After Fourier transform, we find that $H^\mu_n = \sum_{{\bf k}, \eta = y,z} (A^\mu_\eta)^\dagger \mathcal{H}^\mu_{n,\eta} A^\mu_\eta $, with
\begin{eqnarray}
\mathcal{H}^\mu_{n,\eta} = \begin{pmatrix}
\alpha^\mu_{n,\eta} & \gamma^\mu_{n,\eta} \\
(\gamma^\mu_{n,\eta})^\dagger & \alpha^\mu_{n,\eta}
\end{pmatrix},
\end{eqnarray}
where $\alpha^\mu_{n,\eta}$ and $\gamma^\mu_{n,\eta}$ are $16\times16$ Hermitian matrices. The $H^\mu_n$ can be straightforwardly diagonalized to get the energies of BC and PL, which can be used to estimate the finite stiffness regimes of BC and PL shown in the main texts. Below we list the matrix elements for $\alpha^\mu$ and $\gamma^\mu$ in BC and PL respectively combining the contributions from nearest-neighbor, second-neighbor, and third-neighbor terms (i.e., $\alpha=\sum_n \alpha_n$, $\gamma = \sum_n \gamma_n$):
\begin{eqnarray}
&& \alpha^{BC}_{y,a~a=1\sim16} = \frac{9}{2}J_3 - 2 K_3, \\
&& \alpha^{BC}_{y,1~4} = \alpha^{BC}_{y,9~12}= \frac{K_1}{2} e^{-i k_y},\\
\nonumber && \alpha^{BC}_{y,1~5} = \alpha^{BC}_{y,2~3}= \alpha^{BC}_{y,3~7}= \alpha^{BC}_{y,5~6}= \alpha^{BC}_{y,6~10}=\alpha^{BC}_{y,7~8}=  \alpha^{BC}_{y,8~12}=\alpha^{BC}_{y,9~13}=  \alpha^{BC}_{y,10~11}= \alpha^{BC}_{y,11~15}= \alpha^{BC}_{y,13~14}=\alpha^{BC}_{y,15~16}=  \frac{K_1}{2},\\ \\
&& \alpha^{BC}_{y,1~6} = \alpha^{BC}_{y,2~7}=\alpha^{BC}_{y,3~8}= \alpha^{BC}_{y,5~10}= \alpha^{BC}_{y,6~11}=  \alpha^{BC}_{y,7~12}= \alpha^{BC}_{y,9~14}= \alpha^{BC}_{y,10~15}= \alpha^{BC}_{y,11~16}=  \frac{K_2}{2},\\
&& \alpha^{BC}_{y,1~16} = \frac{1}{2}K_2 e^{-i (k_x + k_y)}, \\
&& \alpha^{BC}_{y,2~14}= \alpha^{BC}_{y,4~16}= \frac{K_1}{2} e^{-i k_x},\\
&& \alpha^{BC}_{y,2~13}= \alpha^{BC}_{y,3~14}= \alpha^{BC}_{y,4~15}= \frac{K_2}{2} e^{-i k_x},\\
&& \alpha^{BC}_{y,4~5}= \alpha^{BC}_{y,8~9}= \alpha^{BC}_{y,12~13}= \frac{K_2}{2} e^{i k_y},  
\end{eqnarray}
where $\alpha^{BC}_{y, ab} = (\alpha^{BC}_{y, ba})^*$ and rest of the elements are zero. Similarly,
\begin{eqnarray}
&& \gamma^{BC}_{y,1~2} =  \gamma^{BC}_{y,2~6}=  \gamma^{BC}_{y,3~4} = \gamma^{BC}_{y,4~8} =  \gamma^{BC}_{y,5~9}=  \gamma^{BC}_{y, 6~7}=  \gamma^{BC}_{y,7~11} =  \gamma^{BC}_{y, 9~10} =  \gamma^{BC}_{y,10~14}=  \gamma^{BC}_{y, 11~12} =  \gamma^{BC}_{y, 12~16}=  \gamma^{BC}_{y,14~15}=\frac{K_1}{2},~~~\\
&&  \gamma^{BC}_{y, 1~3} = \gamma^{BC}_{y,2~4}=\gamma^{BC}_{y,5~7} = \gamma^{BC}_{y, 6~8} = \gamma^{BC}_{y,9~11} = \gamma^{BC}_{y,10~12}=\gamma^{BC}_{y,13~15} = \gamma^{BC}_{y,14~16} = \frac{K_3}{2} \left( 1 + e^{-i k_y}\right),\\
&& \gamma^{BC}_{y,1~8} = \gamma^{BC}_{y,5~12}= \gamma^{BC}_{y,9~16}=\frac{K_2}{2} e^{-i k_y},\\
&& \gamma^{BC}_{y,1~9} = \gamma^{BC}_{y,2~10} = \gamma^{BC}_{y,3~11} = \gamma^{BC}_{y,4~12} = \gamma^{BC}_{y,5~13} = \gamma^{BC}_{y,6~14} = \gamma^{BC}_{y,7~15} = \gamma^{BC}_{y,8~16} = \frac{K_3}{2}\left(1 + e^{-i k_x} \right),\\
&& \gamma^{BC}_{y,1~13} = \gamma^{BC}_{y,3~15} = \frac{K_1}{2} e^{-i k_x},\\
&& \gamma^{BC}_{y,1~14} = \gamma^{BC}_{y,2~15} = \gamma^{BC}_{y,3~16} = \frac{K_2}{2} e^{-i k_x},\\
&& \gamma^{BC}_{y,2~5} = \gamma^{BC}_{y,3~6} = \gamma^{BC}_{y,4~7} = \gamma^{BC}_{y,6~9} = \gamma^{BC}_{y,7~10} = \gamma^{BC}_{y,8~11}=\gamma^{BC}_{y,10~13} = \gamma^{BC}_{y,11~14}=\gamma^{BC}_{y,12~15}=\frac{K_2}{2},\\
&& \gamma^{BC}_{y,4~13} = \frac{K_2}{2}e^{-i (k_x - k_y)},\\
&& \gamma^{BC}_{y, 5~8} = \gamma^{BC}_{y,13~16} = \frac{K_1}{2}e^{-i k_y},
\end{eqnarray}
where $\gamma^{BC}_{y, ab} = (\alpha^{BC}_{y, ba})^*$ and rest of the elements are zero.

The diagonal matrix elements of $\alpha^{BC}_z$ for the $z$-flavored bosons in the BC are $\alpha^{BC}_{z,a~a=1\sim16} = 2J_3 - K_1 - K_2 - 2 K_3$, while the off-diagonal matrix elements can be obtained from those of $\alpha^{BC}_y$ with $K_n \rightarrow J_n$. In addition, the matrix elements of $\gamma^{BC}_z$ can be obtained from those of $\gamma^{BC}_y$ by replacing $K_n \rightarrow J_n$.

The matrix elements for the PL are,
\begin{eqnarray}
&& \alpha^{PL}_{y,a~a=1\sim16} = \frac{9}{2}J_3 - 2 K_3, \\
\nonumber && \alpha^{PL}_{y,1~2} = \alpha^{PL}_{y,1~5} = \alpha^{PL}_{y,2~6}=\alpha^{PL}_{y,3~4} =\alpha^{PL}_{y,3~7} = \alpha^{PL}_{y,4~8} = \alpha^{PL}_{y,5~6} =\alpha^{PL}_{y,7~8} = \alpha^{PL}_{y,9~10} = \alpha^{PL}_{y,9~13} = \alpha^{PL}_{y,10~14}=\alpha^{PL}_{y,11~12} =\\
&&\hspace{9cm} =\alpha^{PL}_{y,11~15} = \alpha^{PL}_{y,12~16} = \alpha^{PL}_{y,13~14} = \alpha^{PL}_{y,15~16} =\frac{K_1}{2},\\
&& \alpha^{PL}_{y,1~6} = \alpha^{PL}_{y,2~5} =\alpha^{PL}_{y,3~8} =\alpha^{PL}_{y,4~7} = \alpha^{PL}_{y,6~11} = \alpha^{PL}_{y,7~10} = \alpha^{PL}_{y,9~14} =\alpha^{PL}_{y,10~13} = \alpha^{PL}_{y,11~16} = \alpha^{PL}_{y,12~15} =\frac{K_2}{2},\\
&& \alpha^{PL}_{y,1~6} = \frac{K_2}{2}e^{-i(k_x + k_y)},\\
&& \alpha^{PL}_{y,2~15} = \alpha^{PL}_{y,3~14} = \frac{K_2}{2} e^{-i k_x}, \\
&& \alpha^{PL}_{y,4~13} = \frac{K_2}{2} e^{-i(k_x - k_y)},\\
&& \alpha^{PL}_{y,5~12} = \alpha^{PL}_{y,8~9}=\frac{K_2}{2} e^{-i k_y},
\end{eqnarray}
where $\alpha^{PL}_{y, ab} = (\alpha^{PL}_{y, ba})^*$ and rest of the elements are zero. Similarly,
\begin{eqnarray}
&& \gamma^{PL}_{y,1~3} =  \gamma^{PL}_{y,2~4} =  \gamma^{PL}_{y,5~7} =  \gamma^{PL}_{y,6~8} =  \gamma^{PL}_{y,9~11} =  \gamma^{PL}_{y,10~12} =  \gamma^{PL}_{y,13~15} =  \gamma^{PL}_{y,14~16} = \frac{K_3}{2} \left( 1 + e^{-i k_y} \right),\\
&&  \gamma^{PL}_{y,1~4}=  \gamma^{PL}_{y,5~8} =  \gamma^{PL}_{y,9~12} =  \gamma^{PL}_{y,13~16} = \frac{K_1}{2} e^{-i k_y},\\
&&  \gamma^{PL}_{y,1~8} =  \gamma^{PL}_{y,5~4} =  \gamma^{PL}_{y,9~16} =  \gamma^{PL}_{y,13~12} = \frac{K_2}{2} e^{-i k_y}, \\
&&  \gamma^{PL}_{y,1~9} =  \gamma^{PL}_{y,2~10} =  \gamma^{PL}_{y,3~11} =  \gamma^{PL}_{y,4~12} =  \gamma^{PL}_{y,5~13}=  \gamma^{PL}_{y,6~14} =  \gamma^{PL}_{y,7~15}= \gamma^{PL}_{y,8~16} = \frac{K_3}{2} \left( 1 + e^{-i k_x} \right), \\
&&  \gamma^{PL}_{y,1~13} =  \gamma^{PL}_{y,2~14} =  \gamma^{PL}_{y,3~15}= \gamma^{PL}_{y,4~16} = \frac{K_1}{2} e^{-i k_x},\\
&&  \gamma^{PL}_{y,1~14} =  \gamma^{PL}_{y,2~13} =  \gamma^{PL}_{y,3~16}= \gamma^{PL}_{y,4~15} = \frac{K_2}{2} e^{-i k_x},\\
&&  \gamma^{PL}_{y,2~3} =  \gamma^{PL}_{y,5~9}= \gamma^{PL}_{y,6~7} =  \gamma^{PL}_{y,6~10} =  \gamma^{PL}_{y,7~11}= \gamma^{PL}_{y,8~12} = \gamma^{PL}_{y,10~11} = \gamma^{PL}_{y,14~15}=\frac{K_1}{2},\\
&&  \gamma^{PL}_{y,2~7} =  \gamma^{PL}_{y,3~6} =  \gamma^{PL}_{y,5~10}= \gamma^{PL}_{y,6~9} =  \gamma^{PL}_{y,7~12}= \gamma^{PL}_{y,8~11}= \gamma^{PL}_{y,10~15} =  \gamma^{PL}_{y,11~14}=\frac{K_2}{2},
\end{eqnarray}
where $\gamma^{PL}_{y, ab} = (\alpha^{PL}_{y, ba})^*$ and rest of the elements are zero.

The diagonal matrix elements for the $z$-flavored bosons in the PL are $ \alpha^{PL}_{z,a~a=1\sim16} = 2J_3 -K_1-K_2- 2 K_3$, and the off-diagonal matrix elements of $\alpha^{PL}_z$ can be obtained from those of $y$-flavored bosons with $K_n \rightarrow J_n$. The matrix elements of $\gamma^{PL}_z$ can be obtained from $\alpha^{PL}_y$ with $K_n \rightarrow K_n - J_n$.

Diagonalizing the Hamiltonian to extract the $y$ or $z$ bosons dispersions, we can then determine the regimes where the BC or PL become unstable due to the negative stiffness (i.e. complex eigenvalues near the gapless point in the momentum space). For example, we take a line cut in the Fig.~2(a) parallel to $K_2$-axis. Deep inside the BC/PL regime, we see a single linear-$k$ gapless dispersion at ${\bf k} = {\bf 0}$ representing the spin-dipolar- and quadrupolar-wave modes in the present lattice setup. Approaching the boundary, we observe other local minima near the original linear-$k$ gapless point. At the boundary, we observe that the there are multiple minima in the boson dispersions, which implies the original starting point assuming the BC/PL order is not appropriate. Across the boundary, the original minimum point of gapless spin-dipolar- and quandrupolar-wave modes is no longer a global minimum, and the dispersions at certain momenta ${\bf k}$ become complex, indicating the assumption of the stable BC/PL orders is no longer appropriate, and this suggest the negative stiffness of BC/PL orders.

We then arrive at the conclusions that the BC and PL have different stiffness, and BC and PL can be stable in different parameter regime. In the regime where both BC and PL have positive stiffness, the energies of BC and PL are very close to each other, the difference being $0.1\%$ to $1 \%$, which implies that quantum fluctuations in the $J$-$K$ only model \textit{cannot} efficiently split the degeneracy of BC and PL, consistent with unbiased density matrix renormalization group analysis, and additional inputs such as ring exchanges are needed.

For testing the stable regime of BC/PL order in the $J$-only model, we focus on the Fig.~2(a) in the main texts at fixed $J_2=0.8, J_3=1, K_2=0$, and we vary $J_1$ and $K$. The result is illustrated in Fig.~\ref{Fig:app_J1-K_boundary}. We find that at $K=0$, the BC and PL are both stable only up to a very small threshold, $J_1 = 0.03$. Increasing the value of $K$ substantially increase the stable regimes of BC and PL, although the stable regimes of BC and PL are different in size indicating the difference stiffness of BC and PL.
\begin{figure}[t] 
   \centering
   \includegraphics[width=1.5 in]{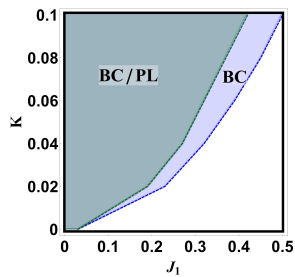} 
   \caption{(Color online)The stable regimes of BC and PL near the purely Heisneberg$J_{n=1,2,3}$ model suggested by flavor-wave theory calculations. At $K=0$, both BC and PL are only stable at very small $J_1 \simeq 0.03$, in which their energies are highly degenerate, the difference being $\sim 0.001\%$. Increasing $K$ substantially increases the stable regime of BC and PL.}
   \label{Fig:app_J1-K_boundary}
\end{figure}

\section{Spin correlation for spin-$3/2$ and the ground-state energy}\label{App:DMRG}

In the main texts, we have shown the spin correlation functions for the spin-$1$ and spin-$2$ models in the (nearly) degenerate BC/PL phase regime obtained from DMRG, where we find the BC state on the TC cylinder and the PL state on the RC cylinder. Here, we also present our DMRG results for the spin-$3/2$ model at the same parameters $(J_1, J_2, J_3, K_1, K_2, K_3)
 = (1.0, 0.8, 1.0, -0.5, 0.1, -0.5)$ in Fig.~\ref{Fig:spin3/2}. The results are consistent those for the spin-$1$ and spin-$2$ systems.

As an additional test of the near degeneracy of the BC and PL states, we compare the ground-state bulk energies of the two states on the different system sizes for the spin-$1$ and spin-$3/2$ models. In Fig.~\ref{Fig:energy}, we demonstrate the energies on the RC cylinder with $L_y = 8$ and the TC cylinder with $L_y = 4, 6$ versus cylinder
width $L$. We note that RC cylinders with $L_y = 4, 6$ are strongly affected by finite-size effects and RC6 does not match the $(\pi/2, \pi/2)$ magnetic ordering. For both models, we find that the energies of the two states on large cylinders are quite close. For the spin-$1$ and spin-$3/2$ models,
the energy differences on the RC ($L_y = 8$) and TC ($L_y = 6$) cylinders are only about $0.2\%$ and $0.1\%$, respectively.

\begin{figure}[t]
    \includegraphics[width= 3.5 in]{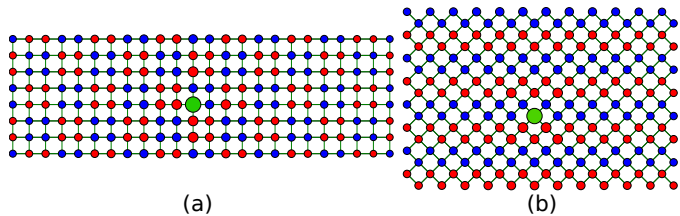}
    \caption{(color online) Real-space spin correlations for the spin-$3/2$ model at $(J_1, J_2, J_3, K_1, K_2, K_3) = (1, 0.8, 1, -0.5, 0.1, -0.5)$ on the RC (a) and TC (b) cylinders with $L_x = 24, L_y = 8$. The green site denotes the reference site in the middle of cylinder. The blue and red circles denote the positive and negative spin correlations of each site with the reference site, respectively. The radius of circle is proportional to the magnitude of spin correlation.}
\label{Fig:spin3/2}
\end{figure}

\begin{figure}[t]
    \includegraphics[width= 3 in]{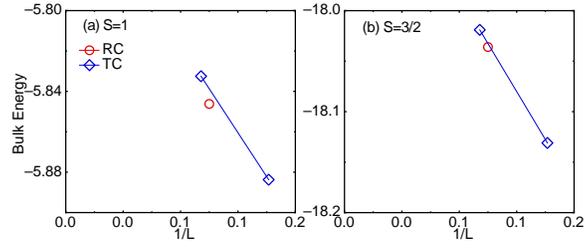}
    \caption{(color online) DMRG ground-state bulk energy versus cylinder width $L$ on both RC and TC cylinders for (a) spin-$1$ and (b) spin-$3/2$ at the parameter $(J_1, J_2, J_3, K_1, K_2, K_3) = (1, 0.8, 1, -0.5, 0.1, -0.5)$. For RC cylinder, the cylinder width $L = L_y$; for TC cylinder, $L = \sqrt{2}L_y$. In the figure, we show the data for the RC cylinders with $L_y = 8$ and the TC cylinders with $L_y = 4, 6$. The energy differences between the RC ($L_y = 8$) and TC ($L_y = 6$) cylinders are only $0.2\%$ and $0.1\%$ for spin-$1$ and spin-$3/2$ models, respectively.}
\label{Fig:energy}
\end{figure}

\end{widetext}
\end{document}